\documentclass[doublecol,figures]{epl2}
\usepackage{graphicx}
\usepackage{amssymb}
\title{Strategy bifurcation and spatial inhomogeneity in a simple
  model of competing sellers}
\shorttitle{Strategy bifurcation in a model of competing sellers}
\author{L. Mitchell\thanks{E-mail: \email{lawrence.mitchell@ed.ac.uk}}
 \and G. J. Ackland\thanks{E-mail: \email{g.j.ackland@ed.ac.uk}}}
\shortauthor{L. Mitchell \etal}
\institute{SUPA, School of Physics, The University of Edinburgh, Mayfield
  Road, Edinburgh EH9 3JZ, United Kingdom}

\abstract{We present a simple one-parameter model for spatially
  localised evolving agents competing for spatially localised
  resources.  The model considers selling agents able to evolve their
  pricing strategy in competition for a fixed market.  Despite its
  simplicity, the model displays extraordinarily rich behaviour.  In
  addition to ``cheap'' sellers pricing to cover their costs,
  ``expensive'' sellers spontaneously appear to exploit short-term
  favourable situations.  These expensive sellers ``speciate'' into
  discrete price bands.  As well as variety in pricing strategy, the
  ``cheap'' sellers evolve a strongly correlated spatial structure,
  which in turn creates niches for their expensive competitors.  Thus
  an entire ecosystem of coexisting, discrete, symmetry-breaking
  strategies arises.}

\pacs{89.65.Gh}{Economics; econophysics, financial markets, business and management}
\pacs{89.75.-k}{Complex systems}
\pacs{89.75.Fb}{Structures and organization in complex systems}

\begin{document}
\maketitle

Many economic models of marketplace interactions have been formulated
(e.g., \cite{Varian:1980,Salop:1977,Salop:1982}).
Generally, these systems assume complete information, and no
transaction costs. That is, for interacting buyers and sellers, buyers
may always search the entire space of sellers (possibly with some
search cost) in order to find the best deal.

When competition is between sellers, such games generally have a
stable, zero-profit (Nash) equilibrium, possibly with multiple prices
\cite{Varian:1980,Salop:1982}.  Despite a substantial literature on
spatial extensions to classic game theoretic models such as the
Prisoner's Dilemma
\cite{Nowak:1992,Huberman:1993,Feldman:1993,Nowak:2004a}, few models
exist for simple marketplaces in which buyers cannot access the entire
seller space (some exist for real-world situations, see, e.g.,
\cite{Kirman:2001}).  Again in the context of the Prisoner's Dilemma,
much progress has been made by considering players which are only
adaptive by random mutation with selection
\cite{Axelrod:1987,Axelrod:1988} as in ecological models
\cite{Holland:1992,Dieckmann:2000}.  However, this has been only
infrequently applied to marketplace games, one model is described in
\cite{Nagel:2000}.

In this paper we present a simple spatial model, similar in spirit to
the Minority Game \cite{Challet:1997,Choe:2004}, with limited interaction
distances and random mutations with selection.  The model is
formulated in terms of active, evolving sellers competing for passive
buyers.  A dual ecological model involves different species competing
for a scarce resource.

\begin{figure}
  \onefigure[width=80mm]{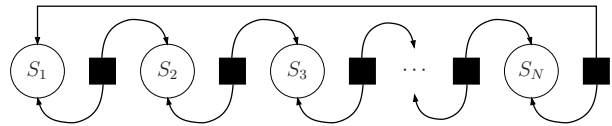}
  \caption{Diagram of buyer-seller connections in 1D.  Buyers and
    sellers are shown by black squares and open circles respectively.
    Arrows indicate the sellers an individual buyer may
    visit\label{fig:figure1}}
\end{figure}

We attempt to make the simplest possible model for a spatially
distributed market with localised information and evolving price
strategy.  We consider a system of $2N$ interacting agents: agents are
split into one of two types, there are $N$ selling agents (sellers)
and $N$ buying agents (buyers).  Agents are placed on a 1-dimensional
chain where each site contains a seller, and each link a buyer. Buyers
are connected to their nearest neighbours, i.e., they have knowledge
of 2 sellers (fig.~\ref{fig:figure1}.  Each seller has capital $C_i$
and an unvarying price $P_i$.  Initial prices are drawn from $P_i \in
[1,P_{max}]$.

Each iteration proceeds as follows:
\begin{enumerate}
\item All sellers' capital is reduced by 2, the cost of producing
  enough stock for both possible buyers.
\item Each buyer visits the cheapest connected seller.
\item For each buyer visiting seller $i$, $C_i$ increases by $P_i$.
\item All sellers with $C_i<0$ are bankrupt: site $i$ becomes vacant.
\item Vacant sites are repopulated with probability $\gamma$.
\item New sellers at site $i$ have $C_i=0$.
\item New sellers at site $i$ take the price of an existing seller at
  randomly chosen site $j$, $P_i=P_j+dp$ ($dp
  \in[\max{(-\Delta,1-P_j)}, \Delta]$).
\end{enumerate}

Note that buyers are always present, but unlike in the other games
mentioned above, sellers' sites might not participate in all rounds of
the game (if $\gamma \neq 1$).  This allows for local variation in the
spatial structure and the availability of supply.  Sellers are assumed
to know their overhead cost (2), and will not charge below this.
Stock is assumed to be perishable and thus, any unsold stock is
destroyed\footnote{Alternately, stock could have negligible value
compared to fixed costs}.  The new sellers may be regarded either as
independent sellers adopting their strategy from successful rivals, or
franchises of those rivals.

Similarly, it is a matter of definition whether the sellers are in any
sense ``intelligent''.  A seller makes no price adjustment between its
initial appearance and bankrupcy, so in this sense exhibits no
intelligence.  It may be assumed to have no information about its
neighbours' strategy for the upcoming round, which would in turn
prevent it from deducing an optimal strategy: as we shall see, in the
evolved state there is a high turnover of shops such that two
neighbours seldom compete for more than one round.  The {\it sites},
by contrast, do have a degree of intelligence, since when their
strategy is observed to have made a loss they change it to one which
has been successful elsewhere.  There is strong evidence that independent 
businesses do indeed adopt known successful business plans, or that 
``best practice'' within a franchise spreads from
one location to another.

There are three parameters in the model, $P_{max}$, $\Delta$, and
$\gamma$.  $P_{max}$ is simply a boundary on the initial conditions;
as we shall see, for reasonable values, the mutation step $\Delta$
affects only the timescale of reaching equilibrium: $\gamma$ is the
only parameter which governs system behaviour.

We will show that this model produces very complex behaviour, with a
range of discrete but non-symmetric strategies emerging.  Before doing
so, we discuss what would be expected from a mean field approach.

The classic analysis for this type of demand limited competition
\cite{Bertrand:1883} suggests that prices will be driven
down to the ``Bertrand equilibrium'', a level that recoups the
production cost, here $P_0=1$.  With the current model, there is
insufficient demand to support all sellers at this price, thus there
will be dead sites whose number may be estimated.

Initially, consider the case where the system is already in the
Bertrand equilibrium: let the price of each seller be chosen randomly
from a uniform distribution $P \in [1, 1 + \delta]$, with small
variation: $\delta \sim \Delta \ll 1$.  In order to survive a round,
each seller must sell all its stock.

Consider a live seller, at the beginning of a round it will be in one of
three situations:
\begin{enumerate}
\item Both neighbouring sellers are dead.
\item One neighbouring seller is dead, while the other is alive.
\item Both neighbouring sellers are alive.
\end{enumerate}
Let $\alpha$ be the proportion of live sellers at the beginning of the
round, then we can write the probability of each of the three cases
as: $p_1 = (1-\alpha)^2$, $p_2 = 2\alpha(1-\alpha)$, and $p_3 =
\alpha^2$.  In order to survive, the seller must either be in
situation (1), or in situation (2) or (3) and outcompeting the live
sellers.  This gives a survival probability (given the uniform price
distribution) of $ p_s = p_1 + \frac{p_2}{2} + {p_3}\int^1_0 (1-x)^2
dx = \frac{\alpha^2}{3} - \alpha + 1$.  The proportion surviving is
hence $f(\alpha) = \alpha(\alpha^2/3 - \alpha + 1)$.  Thus, with
$\gamma = \frac{1}{2}$, at the beginning of the next round, the
proportion of live sites is $\frac{1}{2}\left(1 + f(\alpha)\right)$;
in the steady state, this must be equal to $\alpha$.  Solving
numerically gives the proportion of live sites in the steady state as
$\alpha_{ss} \approx 0.66$.

An alternative assumption is to search for Nash equilibrium of the
game.  Although our agents are constrained to have fixed price
(so-called pure strategies) it is known that the ensemble average at a
Nash equilibrium of pure strategies is the same as the time-average
for mixed strategies, provided the pure agents do not know which
strategy they are playing against \cite{Smith:1982}.  Thus we might
guess that out distribution of prices will resemble the mixed Nash
equilibrium for the non-spatial game.

In this  analysis, we  assume that there  is a distribution  of prices
$f(p)$ which  includes the Bertrand price $P_0$.   In Nash equilibrium
with mixed strategies, all  strategies have the  same payoff  - since
$P_0$ has  zero payoff, other  strategies which offer zero  payoff are
included.  Since all prices below  $P_0$ always lose, we need consider
only higher prices:
\begin{equation}
  \int_{p_i}^\infty (p_i-1)f(p)dp =  \int_1^{p_i} f(p)dp
\end{equation}
whence $f(p)=1/p^2$.  This approach ignores the possibility that sites
are unoccupied.  We may include this in the analysis by allowing an
unoccupied site, paying no overhead, to be part of the strategy (it
has the same payoff as $P_0$).  It turns out, however, that the mixed
strategy equilibrium does not contain this particular pure strategy:
let the probability that a site plays be $\eta$, now suppose that an
opposing site chooses to play with probability $\beta$.  In order to
maximize our expected profit, we should now choose $\eta > \beta$
(cashing in when our opponent plays dead).  Equally, however, our
opponent should choose $\beta > \eta$, to maximize her expected
profit.  Thus, the equilibrium situation is for both players to choose
$\eta = \beta = 1$, i.e., to play every round.

Simulated results with initial prices seeded close to the Bertrand
equilibrium show that the mean field assumption is invalid.  We find
$\alpha = 0.71 \pm 0.01$ in the steady state, which does not agree
with the prediction for $\alpha_{ss}$.  Closer examination of the
structure of the steady state in simulation shows that there is a high
degree of correlation in placement of sellers.  If the steady-state
were a mean field, we would expect $p(n) \approx \alpha_{ss} \;
\forall n$.  As can be seen in fig.~\ref{fig:figure3}, this is
evidently not the case.  An ordered array of ``supercheap'' sellers on
alternate sites forms with prices very close to $P_0$: $P_i-P_0 \ll
\Delta$.  The presence of such an array is stable against intrusion in
the intermediate sites, as a putative new seller opening there must be
cheaper than \textit{both} neighbours, and \textit{both} their
eventual replacements to survive.

The fully correlated case, where every other seller is supercheap, has
$\alpha_{ss} = 0.75$, while the uncorrelated case has $\alpha_{ss}
\approx 0.66$.  In between the two extremes is the actual situation.
With random initial conditions, many correlated regions develop at the
same time.  In order for them to match at their boundaries, they must
nucleate in phase, otherwise they form an antiphase boundary which
cannot be removed by the addition or removal of a single supercheap
seller (fig.~\ref{fig:figure3}).  Thus the Bertrand ``equilibrium'' is
locally stable to small perturbations, although some spatial structure
is already visible (fig.~\ref{fig:figure3}).

\begin{figure}
  \onefigure[width=80mm]{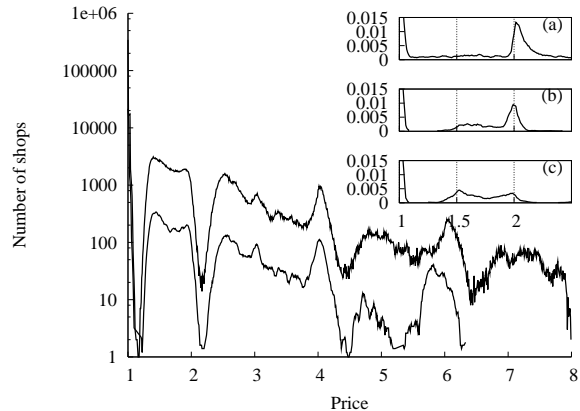}
  \caption{Steady state price distribution for $N=10^5$ and $N=10^6$,
    $\gamma = \frac{1}{2}$, $P_{max} = 8$, $\Delta = 0.04$, price
    shown in units of $P_0$, y axis is absolute number of shops.
    Insets shows evolution of the $P=2$ band to a steady state for
    $N=10^5$, y axis is fraction of total shops.  Note that main
    figure is a log-linear plot, while insets are linear. (a) is after
    100 timesteps, (b) after 500, and (c) in the steady state.  A
    sharp band forms initially above $P=2$ which creates a niche for
    shops with prices $P<2$, the band then migrates downward and
    broadens to that in (c).  Data binned by rounding to 3 decimal
    places, and subsequently smoothed with a 5 point average.  The
    main features of the graph are size independent and reproducible
    (as shown), and sharpen with reduced
    $\Delta$\label{fig:figure2}}
\end{figure}

Simulation of the model with a wider range of initial prices shows
that the global steady-state is a good deal more complex: a range of
high-price sellers coexist with the cheap ones:
fig.~\ref{fig:figure2}.  These sellers exploit temporary
monopoly situations where adjacent sites are dead.

This extraordinary behaviour is at variance with a conventional
demand-limited picture, and can be likened to biological speciation.
Several distinct seller types emerge, which cannot mutate into one
another.  The expensive sellers need not have an infinite lifetime:
because of the replicator dynamics it suffices that each should be
replicated once in its average lifetime.
\begin{figure}
  \onefigure[width=80mm]{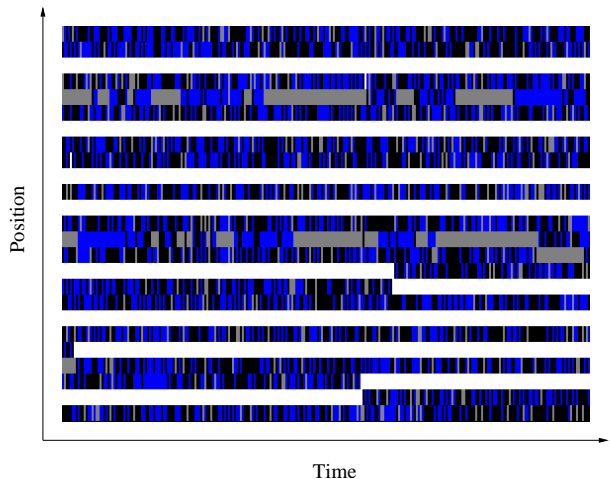}
  \caption{Migration of phase boundaries and correlation amongst cheap
    shops over time.  Each row represents one site (after rebirth),
    coloured according to price.  A dead site is black; a site with
    $P\in [1, 1.00004)$ is white; a site with $P \in [1.00004, 1.008)$
    is grey; a site with $P \in [1.008, 1.28)$ is blue.  Each column
    represents one timestep, the picture shows around $300$ timesteps
    in total\label{fig:figure3}}
\end{figure}

Due to the highly correlated environment which expensive sellers
occupy, a mean field analysis is insufficient.  Instead, consider the
first band ($P=2$) of expensive sellers in 1D: these survive if they
sell to, on average, one buyer per turn.  The possible changes in the
capital $C$ of such an expensive seller, assuming its neighbours are
cheap, are:
\begin{enumerate}
\item $\Delta C = P - 1$ if both neighbouring sellers are dead,
\item $\Delta C = P/2 -1$ if one neighbour is dead and the other is alive,
\item $\Delta C = -1$ if both neighbours are alive.
\end{enumerate}
$C$ therefore carries out a random walk halting when the capital
becomes negative.  The na\"{i}ve guess is that this walk is biased in
favour of the negative step; one might expect that $p_1 =
(1-\alpha)^2$, $p_2 = 2\alpha(1-\alpha)$, and $p_3 = \alpha^2$ with
$\alpha \approx 0.68$.  However, simulation tells us that the mean
lifetime of expensive sellers scales with the lifetime of the game,
our na\"{i}ve guess must therefore be incorrect.

It turns out that the long-lived expensive sellers occupy favourable
niches: the ``supercheap'' sites of the correlated phase, i.e., their
second neighbours are supercheap.  In the limiting case, this means
that their first neighbour competitors are dead 50\% of the time.  This
changes the step direction bias in the random walk described above
since the probabilities primarily depend on $\gamma$ rather than
$\alpha$: the mean lifetime of such a walk with
$\gamma\le\frac{1}{2}$ is infinite.

These niches would appear to allow arbitrarily high prices, and any
seller charging $\geq 2$ to survive.  However, one can apply the ideas
of Bertrand competition to the expensive sellers: on a long enough
timescale they will set up adjacent to one another, and capital will
be transferred to the cheaper seller.

The discrete trading rounds mean that integer prices will have better
short term survival prospects: e.g., a sale to one buyer at $4$ in the
first trading period will ensure survival for two rounds, while $3.9$
will only survive one.  In this scenario with two potential buyers,
the advantage for odd integer price is less: e.g., a price of $3$ has
to sell twice to survive an extra round compared with $2$.  If one
starts with a homogenous distribution of initial prices, this leads to
``speciation'': symmetry breaking in the preferred price band
favouring marginally above integer value,
fig.~\ref{fig:figure2}.  Remarkably, once the
speciation has occured, the character of the competition changes
again.  ``Intraspecies'' competition between sellers in the same price
band becomes critical, and prices below the integer values become
viable, until a balance is reached between intra- and inter-species
competition.

\begin{figure}
  \onefigure[width=80mm]{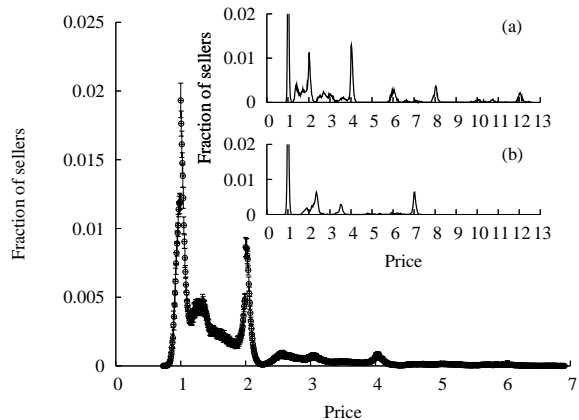}
  \caption{Steady state price distribution for stochastic dynamics
    with $N = 4\times10^4$, $\gamma = \frac{1}{2}$, $\Delta = 0.04$.
    Distribution averaged over final state of 20 ensembles, error bars
    show standard error in mean.  Inset shows steady state price
    distribution for discrete trading rounds and no minimum price for
    connectivities of (a) 4, and (b) 7 buyers per seller (overhead of
    4 and 7 respectively), the peaks at 4 and 7 in (a) and (b)
    correspond to sellers attracting one
    buyer\label{fig:figure4}}
\end{figure}

Although this analysis requires that trading rounds be discrete, the
main features (heavily favoured prices) are still present if trading
happens in a stochastic manner.  For stochastic dynamics, a buyer is
chosen at random to go shopping, and a seller is chosen at random to
pay an overhead, this is repeated such that the expected number of
times a buyer goes shopping is one, this completes one trading round.
The bankruptcy and rebirth dynamics procede as before.  Since buyers
may now visit a seller more than once, there is no upper bound on the
amount of stock a seller may sell, we therefore set the quantity of
stock to $\infty$, and thus the marginal cost to zero.  Despite this,
prices at integer multiples of $P=1$ are still favoured
(fig.~\ref{fig:figure4}).

We may remove a further restriction on the original model by not
requiring that sellers charge above the marginal cost.  In this case,
favoured prices still exist, and do so for a range of connectivities:
for a seller with $d$ potential buyers (paying an overhead $d$),
prices of $P = d n/m,\: n,m \in \mathbb{Z}^+$ are favoured
(fig.~\ref{fig:figure4} inset), corresponding to
attracting $m/n$ buyers on average per round.

We now consider whether such expensive sellers are in some sense
beneficial.  Due to the existence of dead sites and the limited
interaction distance, demand is not completely satisfied.  The
introduction of a wider range of prices results in both a larger total
population and more demand being satisfied.

\begin{figure}
  \onefigure[width=80mm]{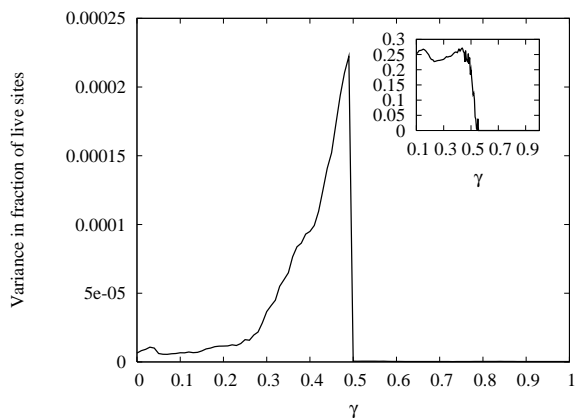}
  \caption{Variance in the time-averaged mean of the fraction of live
    sellers (before rebirth) as a function of $\gamma$, $N=10^4$,
    smoothed using 5 point average separately above and below $\gamma
    = 0.5$.  Inset shows the fraction of expensive shops (after
    rebirth) as a function of $\gamma$\label{fig:figure5}}
\end{figure}

The dependence of the lifetimes of the expensive sellers on $\gamma$
tells us that this parameter may be used to characterize the
distribution.  In the limit $\gamma \rightarrow 1$ all sellers charge
the Bertrand price, as there is is never any chance of expensive
sellers being the only option for consumers.  Equally, $\gamma = 0$
leads to an essentially random distribution of sellers (depending on
initial conditions).  In between these two extremes, we expect some
kind of transition from a regime with expensive sellers to one without
around $\gamma = \frac{1}{2}$: if $\gamma < \frac{1}{2}$ the random
walk of the expensive sellers is biased in favour of the upward,
profit-making step, allowing them to survive indefinitely.

By monitoring fluctuations in a simulation we can see that the system
undergoes a transition at a critical value of $\gamma \approx
\frac{1}{2}$ in which the variance in the number of live sellers
diverges (fig.~\ref{fig:figure5}).

\begin{figure}
  \onefigure[width=80mm]{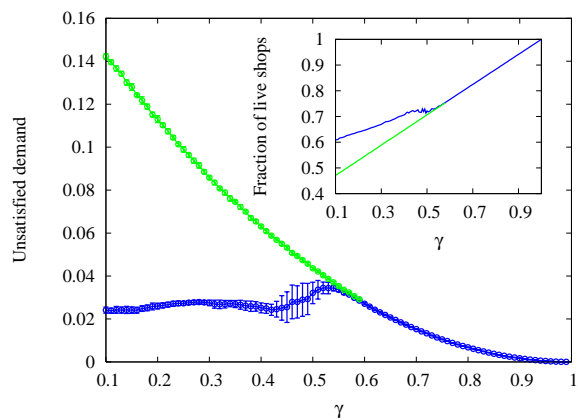}
  \caption{Time-averaged unsatisfied demand as a function of the birth
    rate, $\gamma$, in enforced Bertrand steady state (green) and
    multiple price steady state (blue), $N=10^4$, mean taken as a time
    average over $2\times10^4$ timesteps after the steady state is
    reached.  Inset shows corresponding fraction of live sites after
    rebirth for Bertrand (green) and non-Bertrand (blue) steady state.
    Error bars show standard error in the
    mean\label{fig:figure6}}
\end{figure}

We see further evidence of a transition when considering the mean
unsatisfied demand, being approximately constant for $\gamma <
\frac{1}{2}$ and quadratic for $\gamma > \frac{1}{2}$
(fig.~\ref{fig:figure6}).  If there were no transition, we would
expect the demand to be quadratic in $\gamma$ for all values: this is
indeed the case if we force the system into a Bertrand steady state by
specifying initial conditions accordingly.

The system has a metastable regime for $\gamma < \frac{1}{2}$.  If the
initial conditions only sample the Bertrand regime ($P\approx 1$),
then it remains in such a steady state indefinitely.  Equally, if the
initial conditions sample the whole price spectrum, then the final
steady state contains multiple price bands.  In order for the system
to escape from the Bertrand state, it requires a nucleation of
expensive sellers which cannot happen through mutations in prices (due
to adverse selective pressures on intermediate prices).  Equally, for
small system sizes, fluctuations may eliminate high price bands which
cannot be repopulated.

According to Nash \cite{Nash:1951}, it is possible for multiple
strategies to coexist provided that no individual can do any better by
changing their strategy.  In our system, changes in price at a site
are discontinuous, as are changes in the competing strategies of the
neighbours (each changes only when the shop's capital goes to zero).
These discrete, localised changes prevent the system finding a perfect
Nash equilibrium.  To apply the notion of a Nash equilibrium one has
to assume that the evolution of prices is equivalent to the sites
behaving as rational agents: it is possible that this is the case only
in the infinite time limit, not reached by our simulations.  Another
hypothesis about evolving, replicating systems is that the system as a
whole organises to maximise the number of replicators (here, sellers)
\cite{Ackland:2004}.  In fig.~\ref{fig:figure6} we see good
evidence for this: the expensive sellers become viable when they are
able to increase the total number of living sellers above the Bertrand
solution.  A side effect of this is to minimise the unsatisfied
demand.

We have shown that the obvious generalization of a classic
Bertrand-Edgeworth game has some surprising results.  The classical
Bertrand equilibrium is not necessarily reached, as the system is able
to self-organise to produce niches where different strategies can
flourish.  Further, we have shown that random mutation and selection
can (in the case of restricted initial conditions) produce the
expected Nash result.  That is, sellers need not be active in
selecting strategies, the selective force against badly performing
members is enough to bring the system to equilibrium.

The spontaneous production of evolutionary niches in an initially
homogeneous space has strong parallels in evolutionary ecology.  We
can envisage a similar situation where the ``sellers'' become
individuals foraging for food. The ``cheap sellers'' represent
foragers which are efficient at finding the food, but have a high
metabolic rate and need to feed often.  The ``expensive sellers'' are
less efficient at foraging, but can survive for longer on the same
amount of food.

\acknowledgments
This work was produced by the NANIA collaboration
(\texttt{www.ph.ed.ac.uk/nania}) funded by EPSRC grant T11753.

\end{document}